\documentclass[aps,prb,twocolumn,shortbibliography,superscriptaddress]{revtex4-1}
\usepackage{epsfig}
\usepackage{epstopdf}
\usepackage{amsmath}
\usepackage{amsfonts}
\usepackage{amssymb}
\usepackage{hyperref}
\usepackage{bm}
\usepackage{makecell}
\usepackage{rotating}
\usepackage{hyperref}
\usepackage{multirow}
\usepackage{graphicx}
\usepackage{array}
\usepackage{placeins}
\usepackage{tabularx} 
\newcolumntype{Y}{>{\centering\arraybackslash}X}

\usepackage{graphicx}
\usepackage{dcolumn}
\usepackage{bm}
\usepackage{color}

\usepackage{tikz,xcolor,hyperref}

\definecolor{lime}{HTML}{A6CE39}
\DeclareRobustCommand{\orcidicon}{%
	\begin{tikzpicture}
	\draw[lime, fill=lime] (0,0)
	circle [radius=0.16]
	node[white] {{\fontfamily{qag}\selectfont \tiny ID}};
	\draw[white, fill=white] (-0.0625,0.095)
	circle [radius=0.007];
	\end{tikzpicture}
	\hspace{-2mm}
}

\foreach \x in {A, ..., Z}{%
	\expandafter\xdef\csname orcid\x\endcsname{\noexpand\href{https://orcid.org/\csname orcidauthor\x\endcsname}{\noexpand\orcidicon}}
}

\begin{document}

\title{Relativistic spin-momentum locking in ferromagnets}

\author{Xujia Gong\orcidA}
\email{xgong@magtop.ifpan.edu.pl}
\affiliation{International Research Centre Magtop, Institute of Physics, Polish Academy of Sciences, Aleja Lotnik\'ow 32/46, PL-02668 Warsaw, Poland}

\author{Amar Fakhredine\orcidF}
\email{amarf@ifpan.edu.pl}
\affiliation{Institute of Physics, Polish Academy of Sciences, Aleja Lotnik\'ow 32/46, 02668 Warsaw, Poland}

\author{Carmine Autieri\orcidB}
\email{autieri@magtop.ifpan.edu.pl}
\affiliation{International Research Centre Magtop, Institute of Physics, Polish Academy of Sciences,
Aleja Lotnik\'ow 32/46, PL-02668 Warsaw, Poland}
\affiliation{SPIN-CNR, UOS Salerno, IT-84084 Fisciano (SA), Italy}

\date{\today}
\begin{abstract}
The relativistic spin-momentum locking has been proven in time-reversal-breaking classes of materials with zero net magnetization in the non-relativistic limit, such as altermagnets and other non-collinear magnets. Using density functional theory calculations, we aim to show relativistic spin-momentum locking in ferromagnets, focusing on a broad class of ferromagnetic materials with magnetic sites connected by rotational symmetry, and compare with fcc Ni. In SrRuO$_3$, the antisymmetric exchange interaction produces a spin canting orthogonal to the easy axis, while in all other cases, spin canting is forbidden. Even when the canted magnetic moment in real space is forbidden, relativistic spin-momentum locking shows sizable contributions in k-space. Using prototypical ferromagnets such as orthorhombic SrRuO$_3$, hexagonal CrTe and CrAs with the NiAs crystal structure, half-Heusler MnPtSb, and fcc Ni, we demonstrate that relativistic spin-momentum locking can generate strong effects in ferromagnets. Subdominant components of centrosymmetric ferromagnetic materials with magnetic sites connected by rotational symmetry host spin-momentum locking similar to altermagnets, while noncentrosymmetric MnPtSb hosts relativistic p-wave due to the spin-orbit coupling. Fcc Ni shows a more complex behavior with a combination of two spin-momentum locking patterns characteristic of altermagnets. Because ferromagnets typically have larger bandwidths than altermagnets, they provide a promising platform for observing even-wave relativistic spin-momentum locking and associated emergent phenomena. From an application standpoint, relativistic spin-momentum locking governs symmetry-allowed spin Hall currents, 
and other momentum-dependent spin responses in k-space.
\end{abstract}


\maketitle
	
\section{Introduction}
In altermagnets, the sites with opposite spin are connected by rotational symmetries (proper or improper and symmorphic or nonsymmorphic) but not connected by translation or inversion symmetries\cite{Smejkal22beyond,doi:10.1126/sciadv.aaz8809,hayami2019momentum,hayami2020bottom,Smejkal22,yuan2023degeneracy,Samanta2025,Sun2025,Xu2025,wei2024crystal,Zhang2025,D3NR03681B,D3NR04798A,ssxp-gz9l,Cuono23orbital,D4NR04053H}. While breaking of time-reversal symmetry and weak ferromagnetism induced the spin-orbit coupling were already known for several decades\cite{DZYALOSHINSKY1958241}, one of the most striking novelty of the field of altermagnetism was the non-relativistic spin-momentum locking with even waves for magnetic systems\cite{Smejkal22beyond,song2025unifiedsymmetryclassificationmagnetic}.
Another recent development is the study of multipoles in magnetic systems\cite{doi:10.7566/JPSJ.93.072001,hayami2020bottom,PhysRevB.111.115150}, where the non-relativistic spin–momentum locking of altermagnets corresponds to quadrupole and higher-order multipole structures in k-space in the non-relativistic limit. Beyond their symmetry classification, altermagnets exhibit distinctive nonlinear response phenomena, such as photocurrent generation, which can be employed to probe magnetic structures and magnetoelectric switching\cite{yang2025nonlinear}, as well as ultrafast optical and valley-selective responses\cite{gao2025ultrafast}.

\begin{figure*}
    \centering
    \includegraphics[width=1\linewidth]{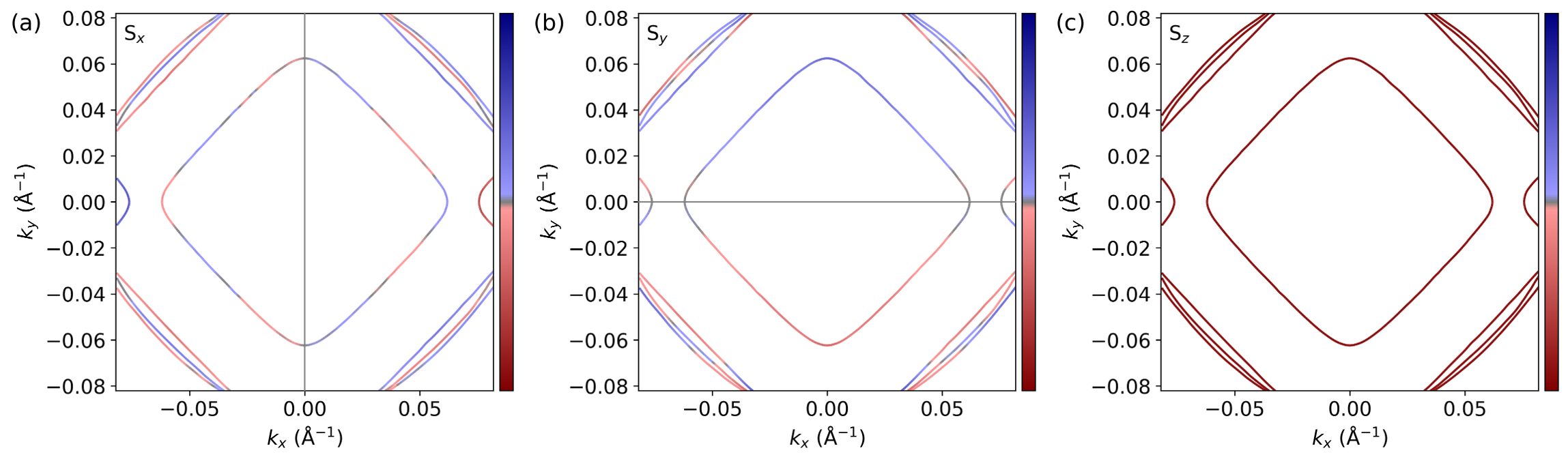}
    \caption{Fermi surface of SrRuO$_3$ with magnetization $||$ $z$-axis for $k_z=0.25$ for (a) the S$_x$ component, (b) the S$_y$ component, (c) the S$_z$ component. Black lines represent the nodal plane for the given spin component. The size of the spin components in the real space is S$_x$ = 0.061 $\mu_B$, S$_y$ = 0.004 $\mu_B$, and S$_z$ = 1.367 $\mu_B$.}
    \label{fig:FermisurfaceSRO_Mz}
\end{figure*}

The spin-orbit coupling preserves the time-reversal symmetry. Therefore, in a system with Kramers degeneracy, the spin-orbit cannot create any magnetism or weak ferromagnetism. In the case of altermagnetic compounds, we have a breaking of the time-reversal symmetry and the spin-orbit could generate the so-called weak ferromagnetism and anomalous Hall effect\cite{PhysRevLett.130.036702,PhysRevB.111.184407}. 
The presence of the altermagnetic spin-splitting is therefore a necessary condition to obtain weak ferromagnetism. In chiral altermagnets, the absence of inversion and mirror symmetries allows spin–orbit coupling to generate persistent spin textures and, in combination with altermagnetic order, can lead to weak ferromagnetism and Néel-vector–dependent spin transport signatures\cite{Tenzin2025,Gauswami2026}.
For historical reasons, the rise of weak ferromagnetism was achieved by introducing the Dzyaloshinskii-Moriya interaction (DMI).
As a consequence of these rotations, these classes of crystal structures host antisymmetric exchange, which can produce weak ferromagnetism depending on the N\'eel vector orientation\cite{839n-rckn}. The simplest and most intuitive form of antisymmetric exchange is the staggered Dzyaloshinskii-Moriya interaction, which produces relativistic weak ferromagnetism in altermagnets orthogonal to the N\'eel vector\cite{PhysRevB.111.054442}. 
The relativistic spin-momentum locking produced is also present in other systems breaking time-reversal symmetry beyond altermagnets, for instance, in the non-collinear MnTe$_2$\cite{Zhu2024}.
When the inversion symmetry is broken, relativistic p-wave spin-momentum lockings can appear if the Rashba or Weyl spin-orbit effect is symmetry allowed in the system\cite{Fakhredine25b,leon2025strainenhancedaltermagnetismca3ru2o7}. Recent studies in perovskite oxide heterostructures have shown that the specific type of spin-orbit interaction—Rashba, Dresselhaus, or a combination—can be directly mapped to the underlying inversion asymmetry of the structure\cite{ganguli2025mapping}. Structural distortions at interfaces or in the bulk selectively generate these spin-orbit couplings, which in turn determine the orientation and nature of the spin-momentum locking.

The materials with spin-momentum locking were proven important for the quantum metrics\cite{doi:10.1126/science.adq3255}. In systems where inversion symmetry is broken, the symmetry allowed Rashba or Dresselhaus spin-orbit interactions provide an additional route to engineer spin-momentum–locked states by structurally controlling the underlying asymmetry\cite{kawano2019designing}.

\begin{figure*}[t!]
\centering
\includegraphics[width=17.9cm,angle=0]{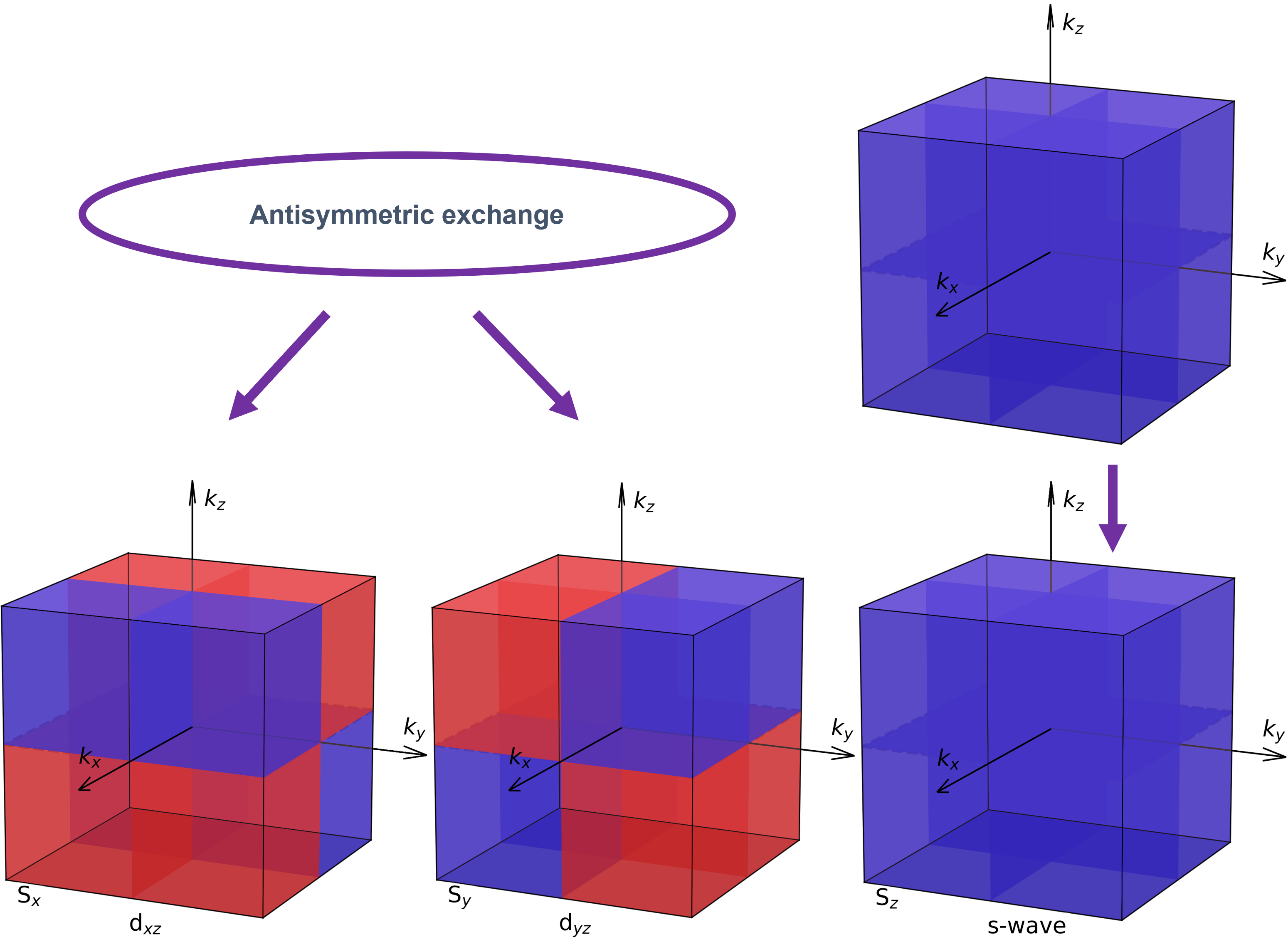}
\caption{Non-relativistic s-wave spin-momentum locking of the ferromagnetic SrRuO$_3$ in the top part. Relativistic Spin-momentum locking for ferromagnetic SrRuO$_3$ with magnetization vector along the $z$-axis in the bottom part. The S$_z$ component is the main component and inherits the s-wave spin-momentum locking from the non-relativistic case. The S$_x$ component is d$_{xz}$-wave , while the S$_y$ component is a d$_{yz}$-wave. The s-wave is represented with a complete Brillouin zone. Red and blue represent regions of the Brillouin zone with opposite spin-splitting.}
\label{fig:RSML_SRO_MZ}
\end{figure*} 

In this paper, we aim to find the even wave spin-momentum locking in the subdominant components of the ferromagnetic phases.
As in altermagnets\cite{AutieriRSML}, we study the spin-momentum locking in ferromagnets, which have the same symmetries as the altermagnets. This material class is the analogue of the altermagnets. However, the difference with respect to simple ferromagnets is limited because both classes exhibit a breaking of time-reversal symmetry.
We aim to study the ferromagnetic phase of compounds with the crystal structures where the spin sites are connected by rotation (proper or improper and symmorphic or nonsymmorphic) but not connected by translation or inversion symmetries. In this class of materials, we choose well-known systems such as the centrosymmetric orthorhombic SrRuO$_3$ and two compounds (CrAs and CrTe) in the hexagonal NiAs structure. We also study the noncentrosymmetric MnPtSb and, finally, we will compare it with the simple ferromagnet Ni in the face-centered cubic structure. We study the  Fermi surface at $k_z$=0 and $k_z$=$\pm$0.25$\frac{\pi}{c}$ where $c$ is the lattice constant, using density functional theory and we aim to evaluate numerically the relativistic spin-momentum locking.

\section{Results}

In each of the following subsections, we will describe one ferromagnetic material using density functional theory calculations. All results reported in this paper include the spin-orbit coupling. The computational framework is described in the Supplementary Materials.

\subsection{Centrosymmetric orthorhombic SrRuO$_3$}

SrRuO$_3$ crystallizes in the space group 62 and experimentally, the easy axis is along the c-vector in bulk crystals\cite{PhysRevB.81.184418,Groenendijk20}. SRO thin films have attracted fundamental research interest due to their intriguing magnetic properties, including the sign change of the anomalous Hall conductivity\cite{Groenendijk20,Vanthiel21}, interfacial charge trapping\cite{lee2022strong} and several other properties. From the structural properties, the four atoms of Ru are perfectly equivalent. Due to the staggered DMI, the four spins are non-collinear.\cite{Benny26} 
The presence of the spin canting in SrRuO$_3$ was already shown using first-principles calculations\cite{PhysRevB.105.245107} and it was confirmed by our spin-resolved density of states reported in supplementary materials.
In this first subsection, we study the relativistic spin-momentum locking of SrRuO$_3$ by considering the magnetization axis along the $x$-, $y$- and $z$-axis.

In Fig. \ref{fig:FermisurfaceSRO_Mz}, we report the Fermi surface at $k_z$=0.25 of SrRuO$_3$ with magnetization along the $z$-axis resolved for S$_x$, S$_y$ and S$_z$ in the panels (a), (b) and (c), respectively. In the figures, we also add the nodal plane for the S$_x$ and S$_y$ components, which are along $k_x$ = 0 and $k_y$ = 0, respectively. 
The size of the canted spin components is much smaller than the dominant one, where S$_x$ is 4\% and S$_y$ is only 0.3\% of S$_z$ in the real space. However, the effect in the k-space is much larger since the spectral weight in the k-space can reach 20\% and 25\% of the dominant component, as shown in the band structure in the supplementary materials.
A schematic representation of the relativistic spin-momentum locking of SrRuO$_3$ with the magnetization along the $z$-axis is reported in Fig. \ref{fig:RSML_SRO_MZ}. A similar calculation at $k_z$= -0.25 was done as well and plotted in Fig. S2, in which S$_x$ and S$_y$ reverse sign with respect to the nodal planes at $k_x$= 0 and $k_y$= 0, respectively, whereas S$_z$ remains unchanged. Combining these data, we obtain that the relativistic spin-momentum locking for the ferromagnetic SrRuO$_3$ with magnetization along the $z$-axis is composed of d$_{xz}$, d$_{yz}$ and s-wave for S$_x$, S$_y$ and S$_z$, respectively. Band structure calculations further validate these results; the corresponding data for the magnetization aligned along the $z$-axis is provided in Figure S7 of the Supplementary Materials. We have reported all the Fermi surfaces at $k_z$= 0 and $k_z$= $\pm$ 0.25 for different magnetization directions in the supplementary materials. A summary of the spin momentum locking for different magnetization directions is reported in Table \ref{tab:SRO}. The notation Q$_0$ stands for the s-wave spin polarization and it seems to always appear for the dominant component at all magnetization directions, as visible in Table \ref{tab:SRO}. Fermi surface calculations for the magnetization along $x$-axis at $k_x$= 0.25 and $k_x$=-0.25 are plotted in Fig. S3 and Fig. S4, respectively, where S$_x$ and S$_y$ preserve their characteristics while S$_z$ flips sign along the nodal plane at $k_x$= 0. Hence, the spin momentum locking are s-wave, d$_{xy}$, and d$_{xz}$ for S$_x$, S$_y$ and S$_z$ respectively. Finally, we show the Fermi surfaces when the magnetization is oriented along the $y$-axis at $k_y$= 0.25 and $k_y$=-0.25 in Fig. S5 and S6, respectively. Comparing the two figures, S$_x$ and S$_y$ do not change while S$_z$ flips sign along 
the nodal plane at $k_y$= 0. Therefore, the spin momentum locking with the magnetization along the $y$-axis is composed by d$_{xy}$, s-wave and d$_{yz}$ for S$_x$, S$_y$ and S$_z$, respectively.
For magnetization along the $x$ and $y$ axes, an additional ferromagnetic component emerges in SrRuO$_3$ that is oriented differently from the dominant magnetization direction. The corresponding $Q_0$ term is small in magnitude and coexists with the magnetic quadrupoles, as reported in Table~\ref{tab:SRO}. This contribution is visible in the band structure shown in Fig.~S38 for the case $\mathbf{M} \parallel y$ at $k_y = 0$, where the magnetic quadrupoles vanish and only the $Q_0$ term remains; since its magnitude is small, it is reported with a small coefficient $\epsilon$ in the Table. Further details on the spontaneous in-plane anomalous Hall effect are provided elsewhere~\cite{Benny26}.

\begin{table}[h!]
\centering
\begin{tabular}{|c|c|c|c|}
\hline
& \multicolumn{3}{c|}{Spin components} \\
\hline
Magnetization direction & $S_x$ & $S_y$ & $S_z$ \\
\hline
$M \parallel x$ & Q$_0$ & Q$_{xy}$+$\epsilon$Q$_0$ & Q$_{xz}$ \\
\hline
$M \parallel y$ & Q$_{xy}$+$\epsilon$Q$_0$ & Q$_0$ & Q$_{yz}$ \\
\hline
$M \parallel z$ & Q$_{xz}$ & Q$_{yz}$ & Q$_0$ \\
\hline
\end{tabular}
\caption{Magnetization directions $M \parallel x$, $y$, and $z$ for orthorhombic SrRuO$_3$ and the corresponding spin--momentum locking for spin components $S_x$, $S_y$, and $S_z$. The coefficient $\epsilon$ is much smaller than one.}
\label{tab:SRO}
\end{table}

\subsection{C\lowercase{r}T\lowercase{e} and C\lowercase{r}A\lowercase{s} with N\lowercase{i}A\lowercase{s} structure}

\begin{figure*}
    \centering
    \includegraphics[width=1\linewidth]{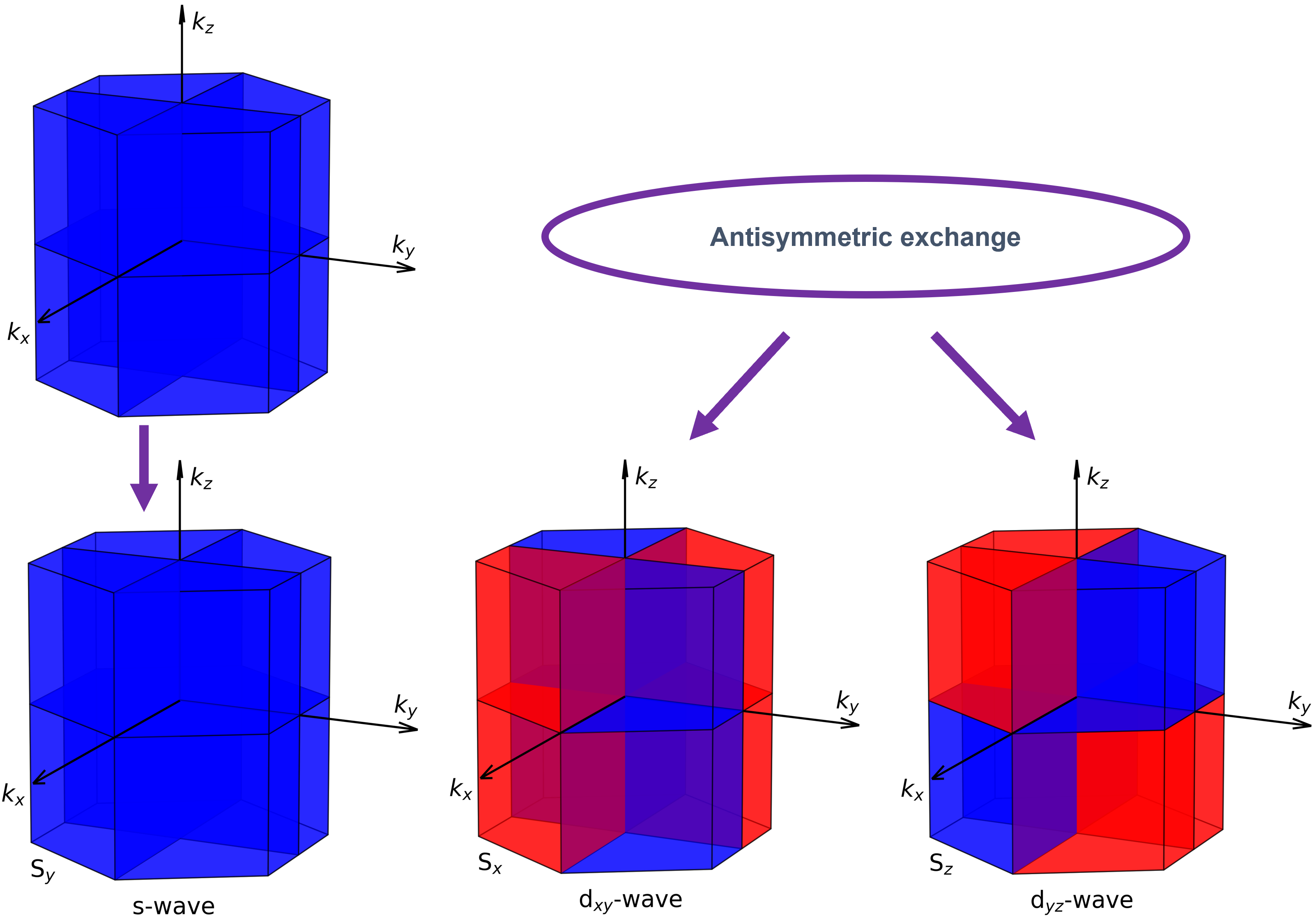}
    \caption{Non-relativistic s-wave spin-momentum locking of the ferromagnetic CrTe in the top part. Relativistic Spin-momentum locking for ferromagnetic CrTe with magnetization vector along the $y$-axis in the bottom part. The S$_y$ component is the main component and inherits the s-wave spin-momentum locking from the non-relativistic case. The S$_x$ component is d$_{xy}$-wave , while the S$_z$ component is a d$_{yz}$-wave. The s-wave is represented with a complete Brillouin zone. Red and blue represent regions of the Brillouin zone with opposite spin-splitting.}
    \label{fig:hexagonal_diagram}
\end{figure*}

CrTe undergoes a paramagnetic-to-ferromagnetic transition at $T_c = 330$~K,  where the ferromagnetic phase exhibits an in-plane magnetic moment. 
Below $100$~K, the system enters an altermagnetic phase with a weak 
ferromagnetism along the $z$-axis~\cite{LUO2025105779}. 
Due to the experimental presence of weak ferromagnetism, the easy axis of magnetization in the 
altermagnetic phase must be along the $y$-axis, and we assume that the easy axis remains along the $y$-axis in the ferromagnetic phase as well.

The magnetic moments of the Cr1 and Cr2 atoms are M$_{Cr1}$=(0,m$_y$,0) M$_{Cr2}$=(0,m$_y$,0), respectively. The size of the magnetic component is m$_y$=4.041, while the other two magnetic components are not allowed by symmetry. This is confirmed by the spin-resolved density of states, which reports no states for S$_x$ and S$_z$ components. Despite the subdominant components being zero in the real space, their spectral weight in the band structure is of the same order of magnitude as the main component, which indicates their significance in the k-space.  We analyze the relativistic spin-momentum locking of CrTe with the magnetization vector along the $y$-axis, which is illustrated in Fig. \ref{fig:hexagonal_diagram}. The spin-momentum locking of the S$_y$ component is s-wave, which it inherits from the non-relativistic part. The S$_x$ component exhibits two nodal planes at $k_x$=0 and $k_y$=0 reflecting a d$_{xy}$ spin-momentum locking, while S$_z$ has d$_{yz}$ wave-like texture where it changes sign along the nodal plane $k_y$=0. The spin-resolved band structures are reported in the supplementary materials in Fig. S16-S18.
Even without spin-canting, CrTe breaks the C$_6$ rotational symmetry, producing the d-wave spin-momentum locking for the subdominant components. This is similar to the relativistic spin-momentum locking of MnTe\cite{AutieriRSML}. To show a case where the magnetization vector breaks the symmetry of the system in a ferromagnet, we study a ferromagnetic system with the NiAs structure (space group no. 194) and a net magnetic moment along the $y$-axis.
The relativistic spin-momentum locking of CrTe with different magnetization vector directions was analyzed and presented in Table \ref{tab:CrTe}. When the magnetization is along the $x$-axis, we find that the spin-momentum locking of the S$_x$ component is s-wave. In contrast, the S$_y$ and S$_z$ components, which arise due to the relativistic part, exhibit a d$_{xy}$ and a d$_{xz}$ spin-momentum locking, respectively. On the other hand, switching the magnetization along the $y$-axis yields an s-wave symmetry for the dominant component S$_y$, d$_{xy}$ for S$_x$, and d$_{yz}$ for S$_z$. Finally, when the magnetization vector is oriented along the $z$-axis, the spin-momentum locking of the S$_z$ component follows an s-wave symmetry reflecting its non-relativistic origin. The S$_x$ component exhibits a d$_{xz}$ spin-momentum locking, while S$_y$ hosts a d$_{yz}$ spin-momentum locking.

\begin{table}[h!]
\centering
\begin{tabular}{|c|c|c|c|}
\hline
& \multicolumn{3}{c|}{Spin components} \\
\hline
Magnetization direction & $S_x$ & $S_y$ & $S_z$ \\
\hline
$M \parallel x$ & $Q_0$ & $Q_{xy}$ & $Q_{xz}$ \\
\hline
$M \parallel y$ & $Q_{xy}$ & $Q_0$ & $Q_{yz}$ \\
\hline
$M \parallel z$ & $Q_{xz}$ & $Q_{yz}$ & $Q_0$ \\
\hline
\end{tabular}
\caption{Magnetization directions $M \parallel x$, $y$, $z$ for CrTe and CrAs with NiAs hexagonal structure and the corresponding spin--momentum locking for spin components $S_x$, $S_y$, and $S_z$.}
\label{tab:CrTe}
\end{table}

\begin{figure*}
    \centering
    \includegraphics[width=1\linewidth]{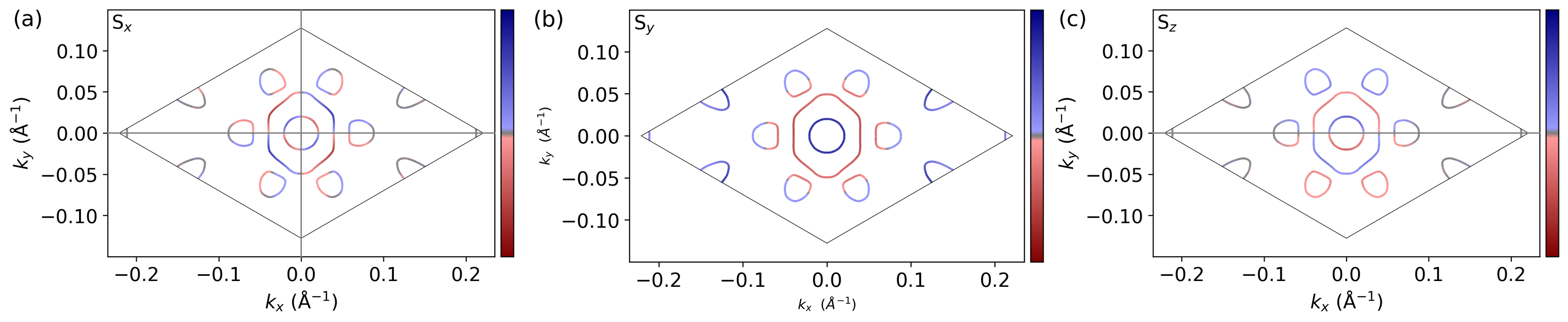}
    \caption{Part of Brillouin zone of the constant 2D energy surface 0.3 eV below the Fermi level of CrTe with magnetization $||$ $y$-axis for $k_z=0.25$ for (a) the S$_x$ component, (b) the S$_y$ component, (c) the S$_z$ component. Black lines represent the nodal plane for the given spin component.}
    \label{fig:crte_kz25_fermi2d}
\end{figure*} 

CrAs has the same spin-momentum locking as CrTe. 
The Fermi surface of CrAs was reported in the Supplementary materials for the three directions of the magnetization at $k_z$=0.25, 0, and -0.25 in Fig. S19-S27.

\begin{table}[h!]
\centering
\begin{tabular}{|c|c|c|c|}
\hline
& \multicolumn{3}{c|}{Spin components} \\
\hline
Mag. & $S_x$ & $S_y$ & $S_z$ \\
\hline
$M \parallel z$ & $Q_{xz} + \alpha Q_{x-y}$ & $Q_{yz} + \alpha Q_{x+y}$ & $Q_0$ \\
\hline
\end{tabular}
\caption{Magnetization directions $M \parallel z$ for noncentrosymmetric MnPtSb and corresponding dominant spin--momentum locking for spin components $S_x$, $S_y$, and $S_z$. Since the system is cubic, the relativistic spin--momentum locking for other magnetization directions can be obtained by rotating the Cartesian axes. The coefficient $\alpha$ highlights that the two terms have different amplitudes.}
\label{tab:MnPtSb}
\end{table}

\begin{figure*}
    \centering
    \includegraphics[width=1\linewidth]{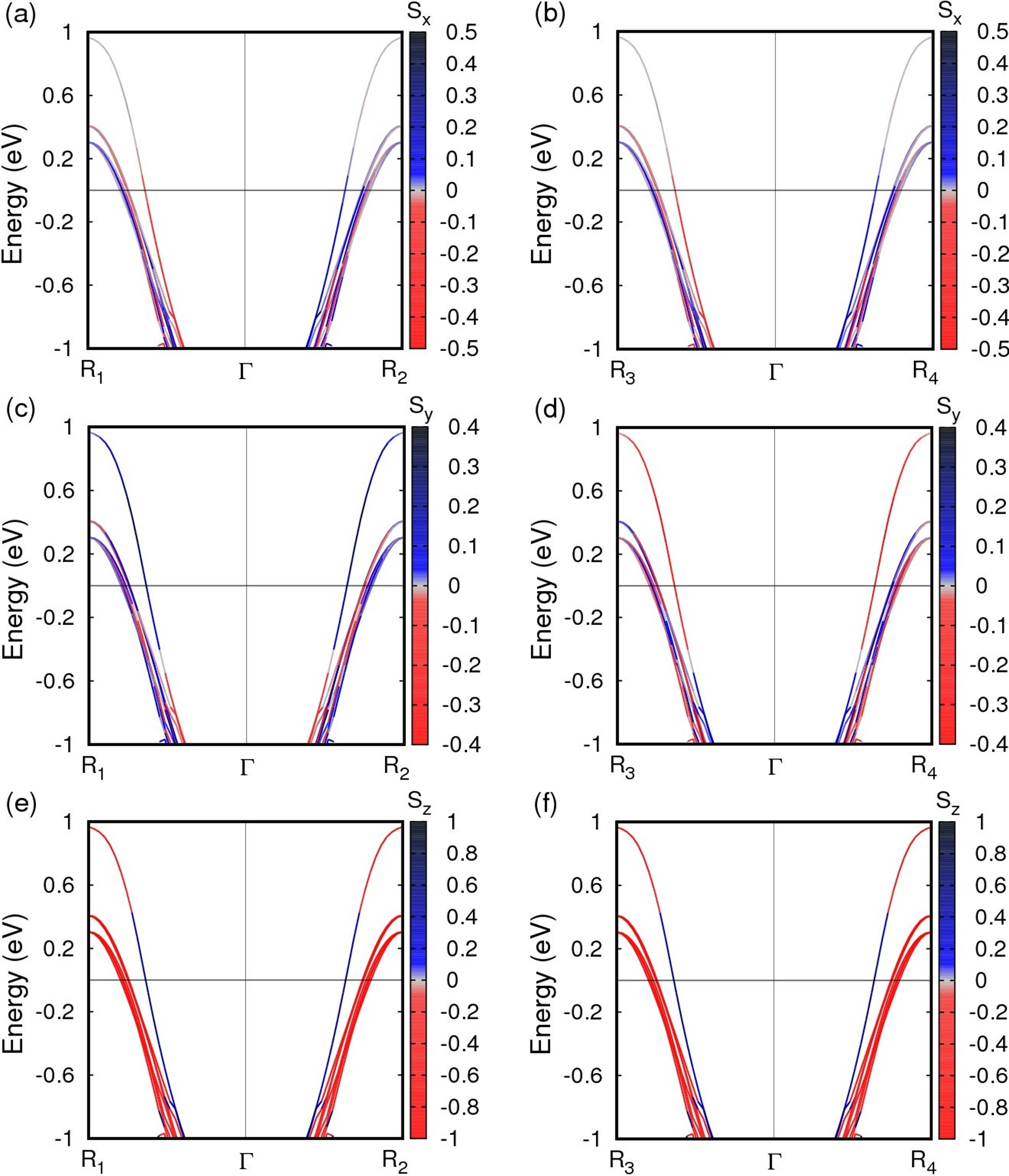}
    \caption{Some of the elements of the relativistic spin-momentum locking for ferromagnetic MnPtSb with magnetization vector along the $z$-axis in the bottom part. The S$_z$ component is the main component and inherits the s-wave spin-momentum locking from the non-relativistic case. The S$_x$ component is p$_x$-p$_y$-wave , while the S$_y$ component is a p$_x$+p$_y$-wave. The s-wave is represented with a complete Brillouin zone. Red and blue represent regions of the Brillouin zone with opposite spin-splitting. For S$_x$ and S$_y$, we also need to add the magnetic quadrupole as reported in the Table \ref{tab:MnPtSb}.}
    \label{fig:mnsbpt_diagram_z}
\end{figure*} 

\begin{table*}[h!]
\centering
\begin{tabular}{|c|c|c|c|}
\hline
& \multicolumn{3}{c|}{Spin components} \\
\hline
Magnetization direction & $S_x$ & $S_y$ & $S_z$ \\
\hline
$M \parallel z$ & $\alpha Q_{(x-y)z} + \beta Q_{(x+y)z}$ & $\alpha Q_{(x-y)z} - \beta Q_{(x+y)z}$ & $Q_0$ \\
\hline
\end{tabular}
\caption{Magnetization directions $M \parallel z$ for fcc Ni and corresponding spin--momentum locking for spin components $S_x$, $S_y$, and $S_z$.  
The coefficients $\alpha$ and $\beta$ highlight the different weights of the multipoles. Since the system is cubic, the relativistic spin-momentum locking for other magnetization directions can be obtained by rotating the Cartesian axes.}
\label{tab:Ni}
\end{table*}

\subsection{Noncentrosymmetric  M\lowercase{n}P\lowercase{t}S\lowercase{b}}

The non-centrosymmetric compound MnPtSb crystallizes in the $F\text{-}43m$ space group \cite{longo2024giant}. In particular, the magnetic phase with zero net magnetization can be identified as altermagnetic. MnPtSb is a member of the half-Heusler alloy family, where only a few ferromagnetic members have been experimentally realized, in contrast to the widely studied full Heuslers\cite{singh2024extended}. Some of these compounds are known to exhibit half-metallic band structures together with very small or fully compensated magnetization, originating from a ferrimagnetic arrangement of the magnetic sublattices \cite{hakimi2013zero,al2021electronic}. The system presents a very strong SOC given by the heavy elements Pt and Sb. When the system is an altermagnet noncentrosymmetric, the magnetic system tends to host Rashba spin-orbit coupling, which is a p-wave relativistic spin-momentum locking\cite{Fakhredine25b,leon2025strainenhancedaltermagnetismca3ru2o7,Cuono26}. 
We first describe our Fermi surface at k$_z$=0 for magnetization direction equal to the z-axis, reported in Fig. S29 of the supplementary materials. The S$_z$ component exhibits an s-wave, while S$_x$ and S$_y$ have nodal lines for k$_x$=k$_y$ and for k$_x$=-k$_y$. Additionally, at k$_x$=0 and k$_y$=0, the spin-splitting is zero. Therefore, we assume that the dominant spin-orbit contribution is a relativistic p-wave of the form p$_x$-p$_y$ for the S$_x$ component, p$_x$+p$_y$ for the S$_y$ component. When we analyze the Fermi surface at k$_z$=$\pm$0.25$\frac{\pi}{a}$ in Fig. S30 and S31, we noticed that for the S$_z$ components the two Fermi surfaces are rotated by 90$^\circ$; therefore, the two Fermi surfaces are not equal due to the breaking of inversion symmetry. Examining the S$_x$ and S$_y$ components, we noted the same sign of the spin in the two regions divided by the plane k$_x$=k$_y$ and k$_x$=-k$_y$ for k$_z$=0,$\pm$0.25$\frac{\pi}{a}$. Therefore, we suppose that the strong spin-orbit splitting dominates in this case. Schematic diagrams of the dominant spin-momentum locking symmetries of the spin components are presented in Fig. \ref{fig:mnsbpt_diagram_z} at a magnetization parallel to the $z$-axis. The S$_x$ component has a p$_x$-p$_y$ symmetry, the S$_y$ component hosts a p$_x$+p$_y$, while the S$_z$ component has the s-wave symmetry, which is inherited from the non-relativistic case before the introduction of SOC. 
To look for the magnetic quadrupoles for the Sx and Sy components, we need to look  
along the diagonal $\Gamma$-R, where the relativistic p-wave spin-momentum locking vanished and we can recover the pure magnetic quadrupole. From the band structure results in Figure S28, we obtain that the spin-momentum lockings for the S$_x$ and S$_y$ components are Q$_{xz}$ and Q$_{yz}$, respectively. Therefore, we can report the relativistic spin-momentum locking in Table \ref{tab:MnPtSb}.
More symmetry analysis is needed to confirm these results for the spin-orbit part.

\subsection{Comparison with the simple ferromagnet Ni fcc}

We also report the comparison between this class of ferromagnets and the simple ferromagnet Ni in the fcc structure. We considered the Ni with magnetization direction along the $c$-axis. We observe an s-wave spin-momentum locking for the dominant and a nodal plane at $k_z$=0 for S$_x$ and S$_y$ components; therefore, all the magnetic quadrupoles present should have a nodal plane at $k_z$=0, which is also confirmed by the sign change of the Fermi surface for $k_z = 0.25$ in Fig.~S33 and for $k_z = -0.25$ in Fig.~S34. The absence of a second nodal plane makes it difficult to determine the exact symmetries of the subdominant component only by density functional theory. The Fermi surface for S$_x$ also exhibits an approximate nodal line for $k_x$=$k_y$, therefore we propose that one quadrupole for S$_x$ is Q$_{(x-y)z}$. Examining the band structure along the $\Gamma$–R direction in Fig.~S32, we find that a single quadrupole is insufficient to explain the results. However, the results can be accounted for by including an additional quadrupole term, Q$_{(x+y)z}$. The results of the relativistic spin-momentum locking are reported in Table \ref{tab:Ni}. These two quadrupoles have different weights, which are accounted for by including appropriate generative coefficients in the table. Even if a deeper explanation of the symmetry of the subdominant components could be determined using an approach with a model Hamiltonian, our DFT results give a clear signature that the Ni fcc host a more complex spin-momentum locking than the ferromagnet with sites connected by rotational symmetry.

\section{Discussion and conclusions}

The similar model Hamiltonian used to describe relativistic spin-momentum locking in altermagnets can also be applied to ferromagnets with antisymmetric exchange. 
We can write the contributions to the non-relativistic Hamiltonian for an effective single-orbital in terms of Pauli matrices for spin and site as:
\begin{align} 
\label{eq:H0skxkz}
\mathcal{H}^0  =& \, \, \, \varepsilon(\boldsymbol{k})\sigma_0^{spin}\sigma_0^{site} \\ 
\nonumber 
\mathcal{H}^{\rm AM}_{S_z}  = & \Delta_z\sigma_z^{spin}\sigma_z^{site} 
\end{align}
For the other two components, the hopping producing the spin-momentum locking is activated by the spin-orbit coupling $\lambda$ via the antisymmetric exchange. We named $\Delta_x$ and $\Delta_y$ the spin-splitting for the relative component and the equations to be 
\begin{equation}
\mathcal{H}^{\rm AM}_{S_x}=\Delta_x\sin(k_x)\sin(k_z)\sigma_y^{spin}\sigma_z^{site}
\end{equation}
and 
\begin{equation}
\mathcal{H}^{\rm AM}_{S_y}=\Delta_y\sin(k_y)\sin(k_z)\sigma_y^{spin}\sigma_z^{site}
\end{equation}
These kinds of model Hamiltonians have been extensively studied for altermagnets\cite{AutieriRSML,PhysRevB.109.024404,PhysRevB.110.144412}, but they can also be used for ferromagnets belonging to the same magnetic space group of altermagnets. The only relevant difference is that in altermagnet the dominant component is an even wave, while in a ferromagnet it is an s-wave. The same spin-orbit terms were proposed to be relevant to the spin dynamics of the ferromagnet SrRuO$_3$\cite{Itoh2016}. 

In this paper, we have demonstrated relativistic spin–momentum locking in several ferromagnets using density functional theory. The relativistic spin–momentum locking of the subdominant components appears regardless of whether spin canting is allowed (as in SrRuO$_3$) or forbidden (as in the other cases considered). Moreover, we have quantitatively estimated the spectral weight of the subdominant components in momentum space. While in real space, the subdominant components are forbidden or orders of magnitude smaller than the dominant component, we have shown that, although smaller, their spectral weight in k-space is of the same order of magnitude as the dominant component.

In SrRuO$_3$, the same staggered DMI that produces the weak ferromagnetism in altermangets can also produce the spin canting in ferromagnets. The spin cantings produce an energy gain in the magnetic energy\cite{AutieriRSML}; more results on the energy gain will be presented elsewhere \cite{Benny26}. The size of the canted spin components in SrRuO$_3$ is 4\% and 0.3\% of the dominant component in the real space for SrRuO$_3$ due to the larger spin-orbit of the magnetic atom. For all other compounds examined in this manuscript, the real-space spin canting is forbidden, while the band structure hosts spectral weights which are of the order of 20\% with respect to the dominant components. This aligns with the trends observed in altermagnets\cite{AutieriRSML}. In orthorhombic ferromagnetic SrRuO$_3$, each of the three spin components exhibits distinct forms of spin–momentum locking with s-wave and two d-waves for every magnetization direction. 
In CrAs and CrTe ferromagnetic phases, we establish that there is no canting for the magnetization vector along the y-axis, while we determine the relativistic spin momentum locking for all magnetization directions. Also in this case, each of the three spin components exhibits distinct forms of spin–momentum locking with s-wave and two d-waves for every magnetization direction. 
In the noncentrosymmetric case, the spin-momentum locking on the three spin-components can host additional p-wave due to Rashba, Dresselhaus or persistent spin-helix being present\cite{Tenzin2025}. MnPtSb is noncentrosymmetric and we propose that the p-wave strongly influences the subdominant components that coexist with magnetic quadrupoles.
We analyzed the case of the simple ferromagnet fcc Ni. In this system, the magnetic atoms are not related by rotational symmetry, and the resulting relativistic spin-momentum locking is more complex. In particular, it cannot be described by a single magnetic quadrupole; instead, two magnetic quadrupoles are required.

The ferromagnets have the same relativistic spin-momentum locking as the altermagnets, except for the relevant fact that the dominant component is s-wave in ferromagnets, while it is even-wave in altermagnets. For instance, the ferromagnetic SrRuO$_3$ with net magnetization oriented along the $z$-axis has the same relativistic spin-momentum locking of G-type YVO$_3$ with N\'eel vector along the $x$-axis\cite{AutieriRSML}. Therefore, we prove that the only difference between altermagnets with weak ferromagnetism and ferromagnets is the size of the net magnetization.

The system with spin-momentum locking can host Nonlinear Photocurrent\cite{doi:10.1021/acsnano.5c01421}.
Moreover, they host octupoles and therefore are a platform for multipolar anisotropy in the anomalous Hall effect\cite{liu2024multipolaranisotropyanomaloushall}.
Our results offer an opportunity to study multipoles and quantum metrics in ferromagnets.
This class of materials could exhibit non-linear Hall effect or circular photogalvanic effect\cite{yoshida2025quantizationspincircularphotogalvanic}.
The magnetic field can be used to change the spin-momentum locking. For instance, if we change the magnetization direction from the $y$-axis to the $x$-axis, we can tune the spin-momentum locking of the S$_z$ component from d$_{yz}$ to d$_{xz}$. From an application perspective, the relativistic spin–momentum locking determines the allowed spin Hall currents\cite{hirakida2025multipoleanalysisspincurrents}, 
and other spin-dependent phenomena as a function of momentum in k-space. 
The relativistic spin-momentum locking of both altermagnet and ferromagnets is a fundamental property of the compounds and should be included in the basic studies of materials and cataloged as the electronic, topological\cite{Bradlyn2017} and other magnetic properties\cite{Jain2013}.

\section*{Acknowledgments}

The authors thank M. Benny, P. Barone and G. Cuono for useful discussions.
This research was supported by the "MagTop" project (FENG.02.01-IP.05-0028/23) carried out within the "International Research
Agendas" programme of the Foundation for Polish Science, co-financed by the
European Union under the European Funds for Smart Economy 2021-2027 (FENG). C.A. acknowledges support from PNRR MUR project PE0000023-NQSTI. We further acknowledge access to the computing facilities of the Interdisciplinary Center of Modeling at the University of Warsaw, Grant g91-1418, g91-1419, g96-1808, g96-1809 and g103-2540 for the availability of high-performance computing resources and support. We acknowledge the access to the computing facilities of the Poznan Supercomputing and Networking Center, Grants No. pl0267-01, pl0365-01 and pl0471-01.

\renewcommand{\bibsection}{\section*{References}}
\bibliographystyle{apsrev4-1}
\bibliography{references}
\end{document}